\documentclass[american,english,jcp,twocolumn]{revtex4-1}
\usepackage[T1]{fontenc}
\usepackage[latin9]{inputenc}
\usepackage{color}
\usepackage{array}
\usepackage{float}
\usepackage{bm}
\usepackage{multirow}
\usepackage{amsmath}
\usepackage{amssymb}
\usepackage{graphicx}

\makeatletter

\@ifundefined{textcolor}{}
{%
 \definecolor{BLUE}{rgb}{0,0,1}
}
\usepackage[colorlinks=true,linkcolor={BLUE},citecolor={BLUE}, urlcolor={BLUE}]{hyperref}

\makeatother

\usepackage{babel}
\begin{document}

\title{Filament actuation by an active colloid at low Reynolds number}

\author{Abhrajit Laskar}
\email{abhra@imsc.res.in}

\affiliation{The Institute of Mathematical Sciences, CIT Campus, Chennai 600113,
India}

\author{R. Adhikari}
\email{rjoy@imsc.res.in}

\affiliation{The Institute of Mathematical Sciences, CIT Campus, Chennai 600113,
India}
\begin{abstract}
Active colloids and externally actuated semi-flexible filaments provide
basic building blocks for designing autonomously motile micro-machines.
Here, we show that a passive semi-flexible filament can be actuated
and transported by attaching an active colloid to its terminus. We
study the dynamics of this assembly when it is free, tethered, or
clamped using overdamped equations of motion that explicitly account
for active fluid flow and the forces it mediates. Linear states are
destabilized by buckling instabilities to produce stable states of
non-zero curvature and writhe. We demarcate boundaries of these states
in the two-dimensional parameter space representing dimensionless
measures of polar and apolar activity. Our proposed assembly can be
used as a novel component in the design of micro-machines at low Reynolds
number and to study elastic instabilities driven by ``follower''
forces. 
\end{abstract}
\maketitle

\section{introduction}

Autonomously actuated slender bodies provide the basic building blocks
for super-diffusive transport at both the cellular and extra-cellular
levels. Classic examples are flagella and cilia \cite{brennen1977fluid}.
Flagellar propulsion endows microorganisms and spermatozoa with motility
while ciliary layers have diverse functions including the transport
of mucous and other biological fluids \cite{lyons2006reproductive,enuka2012epithelial}.
Such active transport keeps, for instance, the trachea free of dust
and aids the transfer of ovum to the uterus. Ciliary dysfunction causes
many human diseases \cite{lyons2006reproductive,tilley2014cilia}.
It is conceivable that biomimetic ciliary layers may have therapeutic
uses. Recent experiments have, in fact, been able to transport genetic
material, therapeutic payloads and functionalized groups to a target
using synthetic propulsion systems \cite{orozco2014bubble,gao2014environmental,jurado2015self,singh2015nano,orozco2015micromotors,duan2015synthetic,wang2013small}.
Enhanced mixing in the context of microfluidics has also been realized
and has numerous technological applications \cite{guix2014nano,singh2015nano}.
The design of autonomously motile micro-machines is now a vigorous
field of research involving a close dialogue between experiment and
theory \cite{camalet1999self,golestanian2007designing,masoud2012designing,keaveny2013optimization,williams2014self}.
Recent surveys of the state-of-the-art are provided in \cite{den2013microfluidic,duan2015synthetic,gao2014environmental}.

In this work, we propose and theoretically analyse a novel mechanism
of actuation of a slender body immersed in a viscous fluid in conditions
where inertia is negligible. We show that a passive elastic filament
can be actuated by attaching an active colloid to its terminus. The
dynamics of a such an assembly is unexpectedly rich when the flow
of the surrounding fluid and the forces it mediates are taken into
account. We study the dynamics, when the other terminus is free, tethered
or clamped, as a function of the leading modes of activity of the
colloid. From this, we identify states of filament motion that are
most suited for biomimetic tasks such as propulsion and mixing. 
\begin{figure}[t]
\selectlanguage{american}%
\includegraphics[width=1\columnwidth]{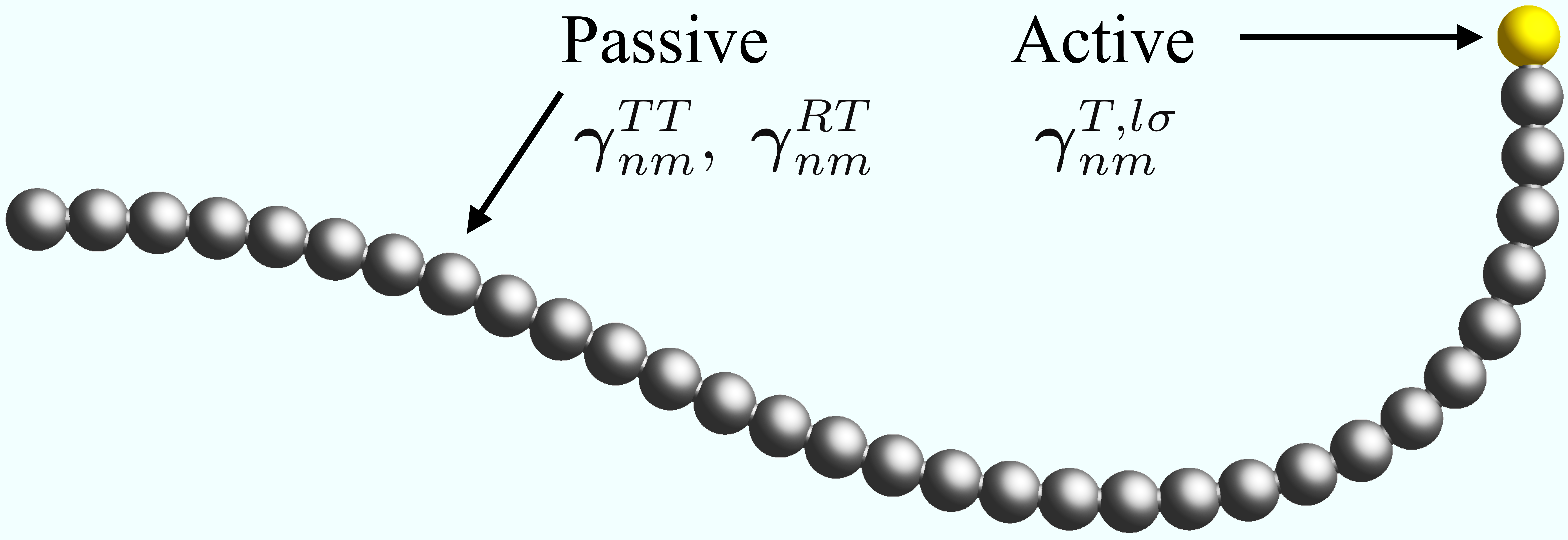}{\caption{Schematic of a passive filament actuated by an active colloid attached
to its terminus. The Stokes and slip friction coefficients indicate
the nature of the hydrodynamic interaction between the passive beads
of the filament and the active sphere (see text). \label{fig:Schema}}
}%
\end{figure}

In recent work, Isele-Holder et al \cite{isele2016dynamics} have
analysed a converse situation, in which an active filament is attached
to a passive colloid. Their study in two-dimensions neglects hydrodynamic
interactions. The importance of hydrodynamic interactions in active
systems is well-established \cite{marchetti2013hydrodynamics}. In
the present context, hydrodynamic interactions can induce instabilities
in filaments that are stable in their absence, and must be considered
in the energy balance that determines the efficiency of active transport.
Our work includes the motion of the ambient fluid in which dynamics
takes place and includes both the hydrodynamic interaction mediated
by the fluid flow and the viscous dissipation that takes place in
the fluid when computing transport efficiencies. 

The theoretical aspect of our study builds on previous work in which
overdamped stochastic equations of motion were derived for the motion
of an active slender body that takes into account its elasticity,
Brownian motion and active fluid flow \cite{laskar2015brownian}.
While the drift terms in the equations of motion were derived systematically
from the solution of the Stokes equations for active spheres, the
diffusion terms were included through a heuristic argument. Here,
we develop the theory of Brownian motion with hydrodynamic interactions
of a \emph{mixture} of active and passive colloids in which \emph{both}
drift and diffusion terms are derived systematically. We recover the
overdamped equations of \cite{laskar2015brownian} when all particles
of the mixture are active. The limit in which exactly one particle
of the mixture is active provides a natural setting for analysing
the dynamics of a passive filament (considered as a chain of passive
particles) actuated by an active colloid, the main focus of this work.

The remainder of the paper is organized as follows. In the Section
\ref{sec:Brownian-microhydrodynamics} we develop the theory of the
hydrodynamically correlated Brownian motion of a mixture of active
and passive spheres. In the Section \ref{sec:Diblock-active-filaments}
following procedure that is now classical \cite{doi1988theory}, we
use the theory of Brownian motion particles to construct equations
of motion of filaments, consisting of a \emph{mixture }of active and
passive ``beads''. In Section \ref{sec:Active-tip} we take the
limit of the above equations in which all but one particle is active
and further restrict the activity to the two modes that produce the
most dominant fluid flow. In Section \ref{sec:Actuating-dynamics}
we perform a thorough numerical study of the resulting equations of
motions and identify states of motion, in the parameter space spanned
by the two dimensionless measures of activity, most suited for biomimetic
applications. A linear stability analysis is performed to locate the
values of these dimensionless measures at which the filament is actuated
and to identify the nature of the dynamical instability through which
this actuation is achieved. We conclude with a discussion and a summary
in the section \ref{sec:Discussions}. 

\section{Brownian microhydrodynamics of active and passive spheres\label{sec:Brownian-microhydrodynamics}}

We consider $M$ passive spheres and $N-M$ active spheres, each of
radius $a$, in an incompressible viscous fluid. The $n$-th sphere
is centered at $\mathbf{R}_{n}$, oriented along $\mathbf{p}_{n}$,
translating with linear velocity $\mathbf{V}_{n}$ and rotating with
angular velocity $\bm{\Omega}_{n}$. The index $n$ runs through $1,\ldots,M,M+1,\ldots,N.$
While the velocity at a point on the boundary of a passive colloid
is its rigid body motion, for an active colloid there is an additional
active slip $\mathbf{v}^{\mathcal{A}}(\boldsymbol{\rho}_{n})$. These
lead to the boundary conditions

\begin{equation}
\mathbf{v}(\mathbf{R}_{n}+\boldsymbol{\rho}_{n})=\begin{cases}
\mathbf{V}_{n}+\boldsymbol{\Omega}_{n}\times\boldsymbol{\rho}_{n} & n\in\mathrm{passive},\\
\mathbf{V}_{n}+\boldsymbol{\Omega}_{n}\times\boldsymbol{\rho}_{n}+\mathbf{v}^{\mathcal{A}}(\boldsymbol{\rho}_{n}) & n\in\mathrm{active},
\end{cases}\label{eq:boundary-conditions}
\end{equation}
which are basic to the mechanics of active and passive spheres. This
method of modelling activity at the surface of the sphere by an additional
velocity slip has been used earlier in the study of electrophoresis
\cite{derjaguin1965molecular,anderson1989colloid,golestanian2007designing}
and in the case of ciliary propulsion of microbes \cite{blake1971spherical}.
The slip in all these studies was assumed to have azimuthal symmetry
about the $\mathbf{p}_{n}$ axis. We differ from these earlier studies
by choosing the most general form of the slip that is consistent with
mass conservation \cite{ghose2014irreducible,singh2014many}. 

The active slip velocity is conveniently parameterised by a series
expansion in tensorial spherical harmonics $\mathbf{Y}^{(l)}$,
\begin{equation}
\mathbf{v}^{\mathcal{A}}\,(\bm{\rho}_{n})=\sum_{l=1}^{\infty}\frac{1}{(l-1)!(2l-3)!!}\mathbf{V}_{n}^{(l)}\cdot\mathbf{Y}^{(l-1)}(\bm{\hat{\rho}}_{n}).\label{eq:slip-expansion}
\end{equation}
Furthermore, the irreducibility of the tensorial harmonics implies
that the expansion coefficients $\mathbf{V}_{n}^{(l)}$ are tensors
of rank $l$, symmetric and irreducible in their last $l-1$ indices.
By using this property, the coefficients can be decomposed into three
irreducible tensors - $\mathbf{V}_{n}^{(ls)},\mathbf{V}_{n}^{(la)},\mathbf{V}_{n}^{(lt)}$
- of rank $l$, $l-1$ and $l-2$; they correspond to \textbf{s}ymmetric
traceless, \textbf{a}ntisymmetric and \textbf{t}race combinations
of the \emph{reducible} indices. We denote the three irreducible parts
by $\mathbf{V}^{(l\sigma)}$ and the suffixes $\sigma=s,a,t$ are
self-explanatory \cite{singh2014many,singh2016traction}. 

For convenience, we introduce special notations for the $l\sigma=1a$
and $l\sigma=2a$ coefficients. These, as will be clear below, are
the linear and angular velocities of an isolated force-free, torque-free
active sphere in an unbounded fluid. They are expressed as $4\pi a^{2}\mathbf{V}_{n}^{\mathcal{A}}=-\int\mathbf{v}^{\mathcal{A}}(\bm{\rho}_{n})dS_{n}$
and $4\pi a^{2}\bm{\Omega}_{n}^{\mathcal{A}}=-\frac{3}{2a^{2}}\int\bm{\rho}_{n}\times\mathbf{v}^{\mathcal{A}}(\bm{\rho}_{n})dS_{n}$.

In the absence of inertia, Newton's equations for the spheres reduce
to instantaneous balance of forces and torques,
\begin{alignat}{1}
 & \mathbf{F}_{n}^{H}+\mathbf{F}_{n}^{B}+\bm{\xi}_{n}^{T}=0,\label{eq:NewtonsLawForce}\\
 & \mathbf{T}_{n}^{H}+\mathbf{T}_{n}^{B}+\bm{\xi}_{n}^{R}=0.\label{eq:NewtonsLawTorque}
\end{alignat}
Here, $\mathbf{F}_{n}^{H}$ and $\mathbf{T}_{n}^{H}$ are the contact
forces and torques applied by the fluid, given by integrals of the
force per unit area $\mathbf{f}=\bm{\sigma}\cdot\mathbf{\hat{\boldsymbol{\rho}}}_{n}$
on the boundary $S_{n}$ of the $n$-th sphere, where $\mathbf{\hat{\boldsymbol{\rho}}}_{n}$
is the local normal and $\bm{\sigma}$ is the Cauchy stress in the
fluid. The spheres may be acted upon by body forces $\mathbf{F}_{n}^{B}$
and body torques $\mathbf{T}_{n}^{B}$, in addition to the Brownian
forces $\boldsymbol{\xi}_{n}^{T}$ and Brownian torques $\boldsymbol{\xi}_{n}^{R}$
due to thermal fluctuations in the fluid. In the absence of activity,
the latter must satisfy a fluctuation-dissipation relation. 

The Cauchy stress that determines the contact forces and torques,
is obtained by solving the fluid mechanical equations of motion. For
an incompressible fluid in slow viscous regime, they are the pair
of Stokes equations $\bm{\nabla}\cdot{\bf v}=0$ and $\bm{\nabla}\cdot\bm{\sigma}=0$,
expressing local conservation of mass and momentum. The Cauchy stress
is $\boldsymbol{\sigma}=-p\boldsymbol{\delta}+\eta(\nabla\mathbf{v}+\nabla\mathbf{v}^{T})$
where $p$ is the pressure and $\eta$ is the viscosity. These pair
of equations must be solved with the boundary conditions on the spheres,
Eq.(\ref{eq:boundary-conditions}), and on any remaining boundaries
of the domain.

By invoking the linearity of the Stokes equations, it can be shown
that the contact forces and torques are linear \emph{functions} of
the boundary condition \cite{singh2016traction}. They are related
to linear velocity $\mathbf{V}_{m}$, angular velocity $\mathbf{\Omega}_{m}$
and irreducible modes of active slip $\mathbf{V}_{m}^{(l\sigma)}$
as,

\begin{alignat*}{1}
{\bf F}_{n}^{H} & =-\boldsymbol{\gamma}_{nm}^{TT}\negthickspace\cdot\mathbf{V}_{m}-\boldsymbol{\gamma}_{nm}^{TR}\negthickspace\cdot\mathbf{\Omega}_{m}-\boldsymbol{\gamma}_{nm}^{(T,\,l\sigma)}\negthickspace\cdot\mathbf{V}_{m}^{(l\sigma)},\\
\text{\ensuremath{{\bf T}_{n}^{H}}} & =\underbrace{-\boldsymbol{\gamma}_{nm}^{RT}\negthickspace\cdot\mathbf{V}_{m}-\boldsymbol{\gamma}_{nm}^{RR}\negthickspace\cdot\mathbf{\Omega}_{m}}_{\mathrm{viscous\,drag}}-\underbrace{\boldsymbol{\gamma}_{nm}^{(R,\,l\sigma)}\negthickspace\cdot\mathbf{V}_{m}^{(l\sigma)}.}_{\mathrm{active\thinspace thrust}}
\end{alignat*}
In the equation above, we have used the summation convention for repeated
particle $(n,m)$ and mode $(l\sigma)$ indices. The first two terms
in each equation are the familiar many-body Stokes drags, expressed
in terms of friction matrices $\boldsymbol{\gamma}_{nm}^{TT}$ , $\boldsymbol{\gamma}_{nm}^{RT}$,
$\boldsymbol{\gamma}_{nm}^{TR}$ , $\boldsymbol{\gamma}_{nm}^{RR}$.
The third terms in each equation are many-body contributions to the
force and torque from activity. The active forces and torques are
$dissipative$ in character but, remarkably, do not vanish when the
spheres are $stationary.$ This reflects the constant consumption
of energy that is necessary to maintain the active slip, independent
of the state of motion of the sphere. A method for computing the $\bm{\gamma}$
matrices in terms of the Green's function has been proposed recently
\cite{singh2016traction}. From there it follows that in an unbounded
fluid, the $\boldsymbol{\gamma}_{nm}^{(T,l\sigma)}$ decay as $r_{nm}^{-l}$
while the $\boldsymbol{\gamma}_{nm}^{(R,l\sigma)}$ decay one power
of distance more rapidly as $r_{nm}^{-(l+1)}$, where $r_{nm}=|\mathbf{R}_{n}-\mathbf{R}_{m}|$
is the distance between the $n$-th and $m$-th spheres. For $n=m$,
all slip friction matrices other than $\boldsymbol{\gamma}_{nn}^{(T,1s)}=6\pi\eta a\boldsymbol{\delta}$
and $\boldsymbol{\gamma}_{nn}^{(R,2a)}=8\pi\eta a^{3}\boldsymbol{\delta}$
are zero. Inserting these in the force and torque balance equations,
with external and Brownian contributions set to zero, shows that $\mathbf{V}_{n}=\mathbf{V}_{n}^{\mathcal{A}}$
and $\boldsymbol{\Omega}_{n}=\boldsymbol{\Omega}_{n}^{\mathcal{A}}$,
justifying their interpretation advertised above. 

The body forces and torques can include externally imposed fields,
interparticle forces, forces with boundaries such as hard walls, and
forces that may arise from constraints such as clamping or pivoting
of the filament. We assume all these forces to be conservative in
character, following, therefore, from the gradient of suitable potentials.

The correlated thermal noises $\bm{\xi}_{m}^{T}$ and $\bm{\xi}_{m}^{R}$
(Eq. \ref{eq:NewtonsLawForce} and Eq. \ref{eq:NewtonsLawTorque}),
obey the fluctuation-dissipation theorem $\langle\boldsymbol{\xi}_{n}^{\alpha}\boldsymbol{\xi}_{m}^{\beta}\rangle=2k_{B}T\,\boldsymbol{\gamma}_{nm}^{\alpha\beta}$,
where $\alpha,\beta=T,R$. We note that there is no compensating source
of fluctuation for the dissipation due to activity, as the latter
arises from non-equilibrium processes that hold the system away from
thermodynamic equilibrium.\textcolor{black}{{} }

The force and torque balance equations can now be used to obtain the
velocity and angular velocity, for given values of slip, body and
thermal contributions to the forces and torques. The equations, though
are implicit in the velocities and angular velocities, which makes
their numerical integration cumbersome. Explicit equations can be
derived by solving the force and torque balance equations for velocities
and angular velocities. In the the study of passive suspensions, this
procedure leads from the ``resistance'' formulation to the ``mobility''
formulation. A similar procedure followed here leads, in addition
to the well-known mobility matrices of Stokes flow, to a new set of
tensorial coefficients that have been named propulsion matrices \cite{singh2014many,laskar2015brownian}.
They are related to the mobility matrices and friction tensors by
\cite{singh2016traction},\begin{subequations}\label{eq:pi-mu-relation}
\begin{alignat}{1}
\bm{\pi}_{nm}^{(\text{T},\,l\sigma)} & =-\left(\boldsymbol{\mu}_{np}^{TT}\cdot\boldsymbol{\gamma}_{pm}^{(T,\,l\sigma)}+\boldsymbol{\mu}_{np}^{TR}\cdot\boldsymbol{\gamma}_{pm}^{(R,\,l\sigma)}\right),\\
\bm{\pi}_{nm}^{(R,\,l\sigma)} & =-\left(\boldsymbol{\mu}_{np}^{RT}\cdot\boldsymbol{\gamma}_{pm}^{(T,\,l\sigma)}+\boldsymbol{\mu}_{np}^{RR}\cdot\boldsymbol{\gamma}_{pm}^{(R,\,l\sigma)}\right).
\end{alignat}
\end{subequations} where $\boldsymbol{\mu}_{nm}^{TT}$, $\boldsymbol{\mu}_{nm}^{RT}$,
$\boldsymbol{\mu}_{nm}^{TR}$ and $\boldsymbol{\mu}_{nm}^{RR}$ are
the usual mobility matrices. We provide explicit expressions for both
the mobility matrices and the propulsion tensors for spheres in an
unbounded fluid in the Appendix. It is convenient to express the correlated
Langevin noises $\bm{\xi}^{T}$ and $\bm{\xi}^{R}$ in terms of uncorrelated
Wiener processes $\bm{\zeta}^{T}$ and $\bm{\zeta}^{R}$ and the Cholesky
factors of the correlation matrix following the usual procedure in
Brownian dynamics. 

With these considerations the equations for Brownian hydrodynamics
of active colloids is 

\begin{widetext}\begin{subequations}

\begin{alignat}{1}
\mathbf{V}_{n} & =\bm{\mu}_{nm}^{TT}\cdot\mathbf{F}_{m}^{B}+\bm{\mu}_{nm}^{TR}\mathbf{\cdot T}_{m}^{B}+\sqrt{2k_{B}T\bm{\mu}_{nm}^{TT}}\cdot\ \bm{\zeta}_{m}^{T}+\sqrt{2k_{B}T\bm{\mu}_{nm}^{TR}}\cdot\ \bm{\zeta}_{m}^{R}+\,\bm{\pi}_{nm}^{(T,\,l\sigma)}\cdot{\bf V}_{m}^{(l\sigma)}\label{eq:flowInTermsofPropulsionMatrix}\\
\bm{\Omega}_{n} & =\underbrace{\bm{\mu}_{nm}^{RT}\mathbf{\cdot F}_{m}^{B}+\bm{\mu}_{nm}^{RR}\cdot\mathbf{T}_{m}^{B}}_{\mathrm{Passive}}+\underbrace{\sqrt{2k_{B}T\bm{\mu}_{nm}^{RT}}\cdot\ \bm{\zeta}_{m}^{T}+\sqrt{2k_{B}T\bm{\mu}_{nm}^{RR}}\cdot\ \bm{\zeta}_{m}^{R}}_{\mathrm{Brownian}}+\,\underbrace{\bm{\pi}_{nm}^{(R,\,l\sigma)}\cdot{\bf V}_{m}^{(l\sigma)}}_{\mathrm{Active}}
\end{alignat}
\end{subequations}\end{widetext}In an earlier paper \cite{laskar2015brownian},
the diffusion terms in this stochastic differential equation were
arrived at by heuristic arguments. Here, we have derived them without
such recourse by appealing directly to force and torque balance, together
with the known form of the Brownian forces and torques acting on particles
in a fluctuating Stokesian fluid. 
\begin{figure*}[t]
\selectlanguage{american}%
\includegraphics[width=0.97\textwidth]{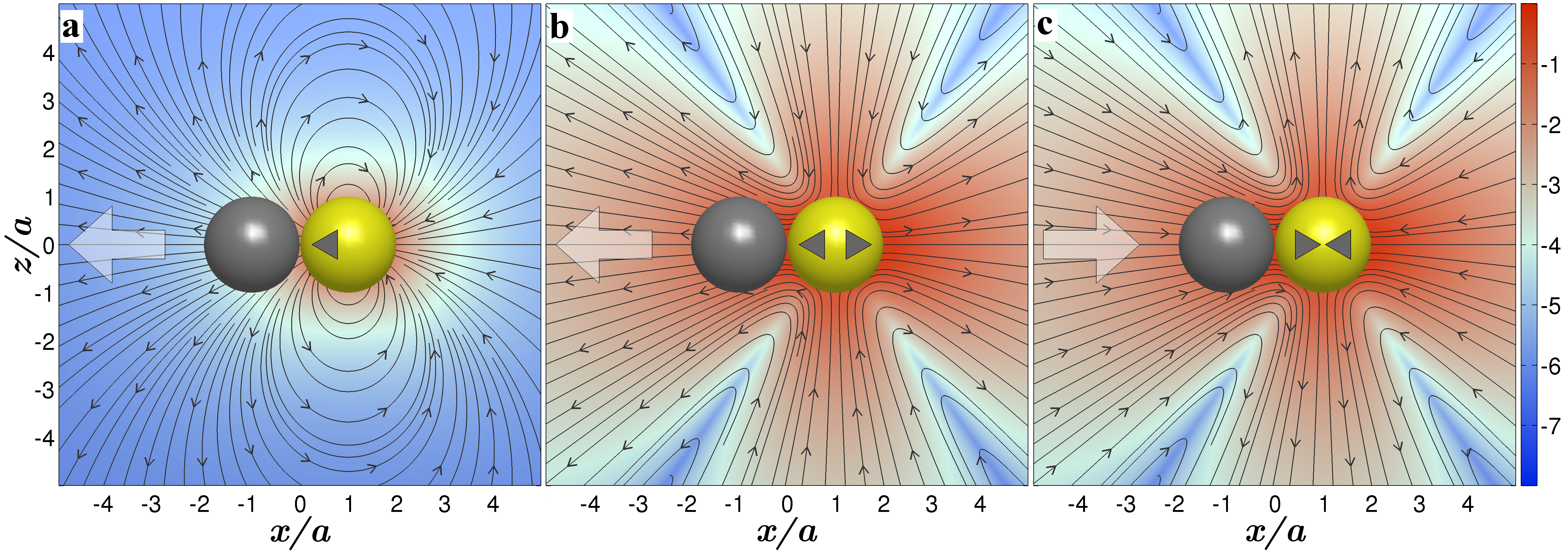}{\caption{Motion of a passive colloid in the flow of an active colloid. The
active is colloid is polar in panel (a), apolar and extensile in panel
(b), and apolar and contractile in panel (c). The streamlines show
active flow in the \emph{absence} of the passive colloid. The background
colour maps to the logarithm of the absolute value of the local flow
field normalised by the global maximum. The large arrows show the
direction of the force on the passive colloid exerted by the active
flow.\label{fig:AtomicFlow}}
}
\end{figure*}

\section{Diblock active filaments\label{sec:Diblock-active-filaments}}

In this section, we use the overdamped equations of motion of an active-passive
mixture presented above to model a diblock active filament, consisting
of chains of active \emph{and }passive beads. In previous work \cite{jayaraman2012autonomous,laskar2013hydrodynamic,laskar2015brownian},
we modelled an active slender body as a chain of active spheres. Here,
we recall key elements of that work, while allowing \emph{some }spheres
to be passive.

A filament is obtained by connecting spheres by a potential $U$ that
exerts forces ${\bf F}_{n}^{B}=-\bm{\nabla}_{n}U$ on the $n$-th
sphere. The potential $U$ is the sum of connective, elastic, and
self-avoiding steric potentials \cite{jayaraman2012autonomous,laskar2013hydrodynamic}.
The relative importance of the $l\sigma$ mode of activity to the
forces due to these potentials is quantified by dimensionless ``activity''
numbers,
\begin{equation}
\mathcal{A}{}^{l\sigma}=\frac{\left|\boldsymbol{\gamma}_{nm}^{(T,\,l\sigma)}\cdot\mathbf{V}_{m}^{(l\sigma)}\right|}{\left|\bm{\nabla}_{n}U\right|}\label{eq:generle-act-num}
\end{equation}
The filament dynamics is sensitive to the orientation of the principal
axes $\mathbf{p}_{n}$ of the velocity coefficients $\mathbf{V}_{n}^{(l\sigma)}$
relative to the local Frenet-Serret frame attached to the filament,
defined by the tangent $\mathbf{t}$, normal $\mathbf{n}$ and binormal
$\mathbf{b}$ vectors. The most general parametrization of the relative
orientation of the principal axis and the local frame is $\mathbf{p}_{n}=\,\alpha_{1}\mathbf{t}_{n}+\alpha_{2}\mathbf{n}_{n}+\alpha_{3}\mathbf{b}_{n}$.
Torsional potentials can be introduced to penalize departures from
this preferred orientation or constraint torques can be used to enforce
the orientation exactly. In what follows, we choose the principal
axis to be parallel to the local tangent and enforce this exactly
through constraint torques. Since the orientation of the spheres is
now subordinated to the local filament conformation, there is no independent
angular degree of freedom. Therefore, the explicit form of the equation
of motion of the diblock active filament is

\begin{eqnarray}
\negthickspace\negthickspace\negthickspace\negthickspace\negthickspace\negthickspace\dot{\mathbf{R}}_{n} & = & \underbrace{\bm{\mu}_{nm}^{TT}\cdot\mathbf{F}_{m}^{B}+\bm{\mu}_{nm}^{TR}\cdot\mathbf{T}_{m}^{B}}_{\mathrm{Passive}}+\underbrace{\bm{\pi}_{nm}^{(T,\,l\sigma)}\cdot{\bf V}_{m}^{(l\sigma)}}_{\mathrm{Active}},\label{eq:EoM-mixture}
\end{eqnarray}
where it is understood that $\mathbf{V}_{n}^{(l\sigma)}=0$ for the
$n=1,\ldots,M$ passive spheres, and that the  torque $\mathbf{T}_{n}^{B}$
contains the constraints needed to maintain the principal axis parallel
to the tangent. \textcolor{black}{An estimate of the relative strengths
of noise and activity shows that activity is between a $100$ to a
$1000$ times more dominant than thermal fluctuations in many typical
situations \cite{wang2013small,singh2016traction}. Accordingly, we
neglect them for the remainder of this work. }
\begin{figure*}
\includegraphics[width=0.97\textwidth]{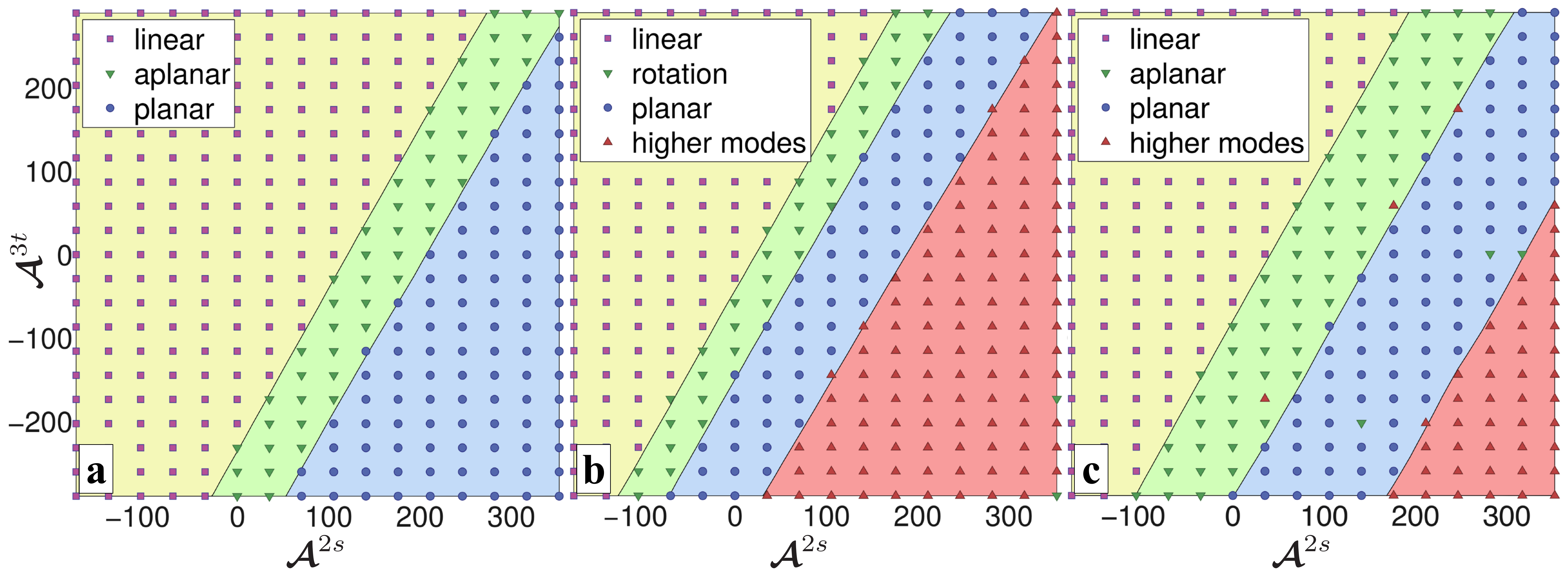}\caption{Stable states of actuation of a filament which is (a) free, (b) tethered,
or (c) clamped, in the parameter space of $\mathcal{A}^{2s}$ and
$\mathcal{A}^{3t}$, the dimensionless measures of apolar and polar
activity. The net active force on the filament is guaranteed to be
compressive in the fourth quadrant, $\mathcal{A}^{2s}>0$, $\mathcal{A}^{3t}<0$.
Each dot represents one simulation and the background colours are
a guide to the eye.\label{fig:PhaseDiag}}
\end{figure*}
\begin{figure*}
\centering{\includegraphics[width=1\textwidth]{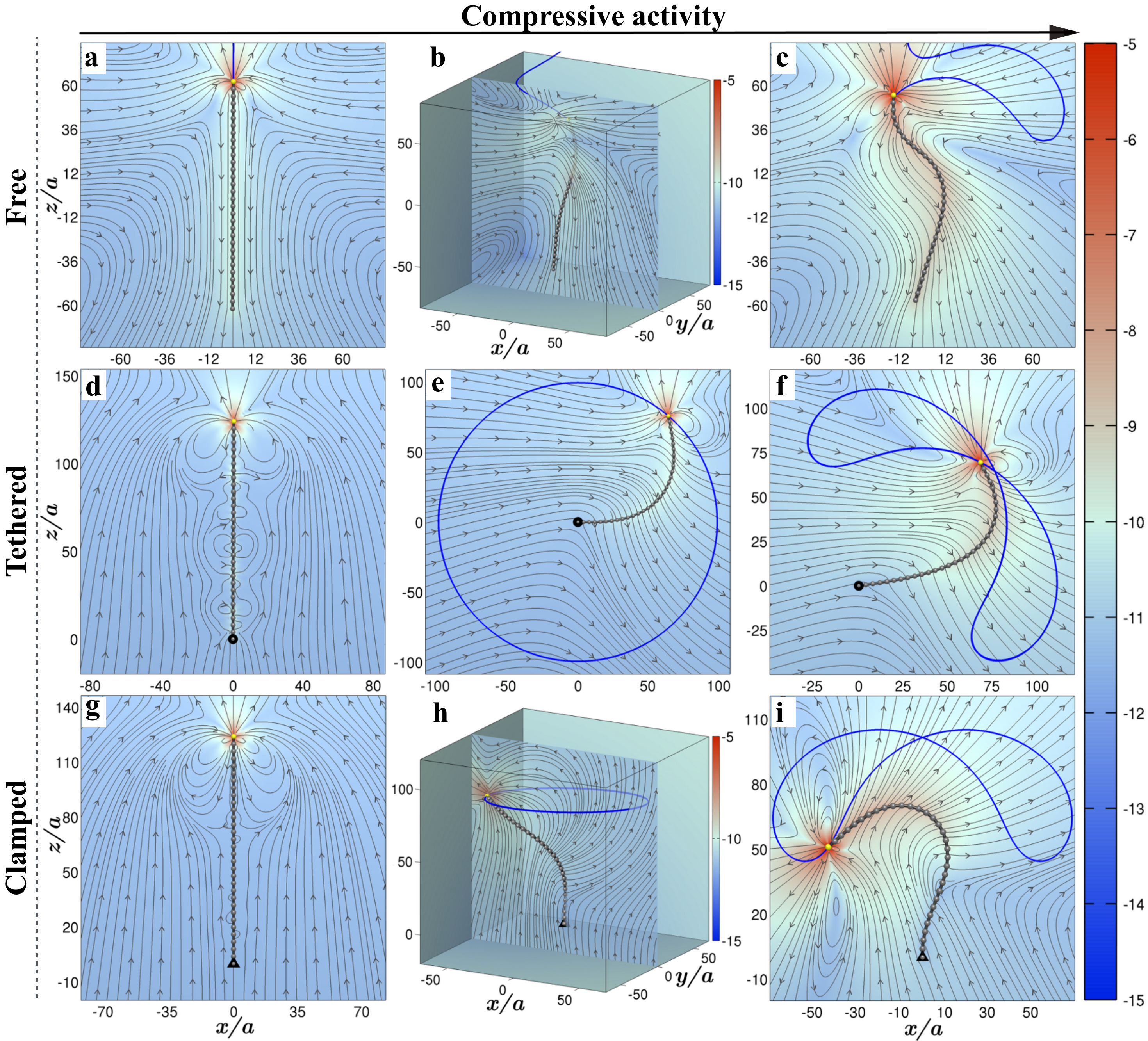}}\caption{States of filament actuation for varying boundary conditions and activity.
The top, middle, and bottom rows show a filament that is free, tethered
and clamped respectively. The activity varies in the $\mathcal{A}^{2s}-\mathcal{A}^{3t}$
in such a manner that the compressive force due to the active colloid
increases from left to right in each row. The precise values are tabulated
in the Appendix. Streamlines show the net flow around the assembly,
with the background coloured as in Fig. (\ref{fig:AtomicFlow}). \label{fig:Flowphase}}
\end{figure*}

We now study a limiting case of these diblock active filaments where
all but one sphere is passive. This provides the model of a passive
filament actuated by an active colloid.

\section{Passive filament - active colloid\label{sec:Active-tip}}

We consider a diblock active filament in which the first $N-1$ spheres
are passive and the $N$-th sphere is active, illustrated schematically
in Fig. (\ref{fig:Schema}). The radii of the active and passive spheres
are allowed to be different. The slip velocity of the active colloid
is truncated at three terms, including the two leading polar terms
and the leading apolar term,
\begin{alignat}{1}
\mathbf{v}(\bm{\rho}_{n}) & =\dot{\mathbf{R}}_{n}-\underbrace{{\bf V}_{n}^{\mathcal{A}}+\tfrac{1}{15}{\bf V}_{n}^{(3t)}\cdot{\bf Y}^{(2)}}_{\mathrm{polar}}+\underbrace{{\bf V}_{n}^{(2s)}\cdot{\bf Y}^{(1)}}_{\mathrm{apolar}}\label{eq:trunc-active-slip}
\end{alignat}
This model is sufficiently general to describe the far-field flow
of a variety of polar and apolar active colloids \cite{ghose2014irreducible}.
We assume that the principal axes of the slip coefficients are parallel
to the tangent vector, $\mathbf{t}_{N}$, at the terminus of the filament,
so that

\begin{gather}
{\bf V}_{N}^{(2s)}=as_{0}(\mathbf{t}_{N}\mathbf{t}_{N}-\frac{1}{3}\bm{\delta})\label{eq:apolar-slip}\\
{\bf V}_{N}^{\mathcal{A}}=-\frac{3}{5}{\bf V}_{N}^{(3t)}=-\frac{3a^{2}}{5}d_{0}\mathbf{t}_{N}.\label{eq:polar-slip}
\end{gather}
 Additionally, we neglect the subdominant contribution from the constant
torques to the equations of motion. With these consideration, the
explicit equation of motion for the filament and the active colloid
are \begin{widetext}\begin{subequations}
\begin{alignat}{1}
\dot{\mathbf{R}}_{n}= & -\frac{1}{6\pi\eta a}\bm{\nabla}_{n}U-\frac{1}{8\pi\eta}\sum_{m\neq n}\mathcal{F}^{0}\mathcal{F}^{0}\mathbf{G}\cdot\bm{\nabla}_{m}U+\frac{7a^{2}}{6}\mathcal{F}^{0}\mathcal{F}^{1}\bm{\nabla}_{N}\mathbf{G}\cdot{\bf V}_{N}^{(2s)}+\frac{a^{3}}{10}\nabla_{N}^{2}\mathbf{G}\cdot{\bf V}_{N}^{(3t)};\;n,m\in\mathrm{filament},\label{eq:EoM-active-bead}\\
\dot{\mathbf{R}}_{N}= & -\frac{1}{6\pi\eta a}\bm{\nabla}_{N}U-\frac{1}{8\pi\eta}\sum_{m\neq N}\mathcal{F}^{0}\mathcal{F}^{0}\mathbf{G}\cdot\bm{\nabla}_{m}U-{\bf V}_{N}^{\mathcal{A}},\qquad\thinspace\thinspace\ \thinspace\thinspace\thinspace\thinspace\thinspace\thinspace\qquad\qquad\qquad\qquad\qquad\:\quad\qquad\mathrm{active\;colloid}.\label{eq:EoM-passive-bead}
\end{alignat}
\end{subequations}\end{widetext} Here, $\mathcal{F}^{l}$ represents
the correction due to the finite size of the spheres over the usual
multipole expansion that implicitly assumes point particles. This
Kirkwood-Riseman pair approximation is know to correct to $\mathcal{O}((a/b)^{3})$,
where $b$ is the mean separation between the spheres of radius $a$
\cite{yoshizaki1980validity,laskar2013hydrodynamic}. 

From the Eq. \ref{eq:generle-act-num}, it is clear that the two activity
modes yield two activity numbers. In dimensionless units, theses numbers
are $\mathcal{A}^{2s}$ and $\mathcal{A}^{3t}$ and expressed in terms
of various system parameters as,
\begin{alignat}{1}
\mathcal{A}^{2s} & =\frac{7\pi\eta a^{4}L^{2}s_{0}}{\kappa b^{2}}\label{eq:apolar-act-num}\\
\mathcal{A}^{3t} & =\frac{18\pi\eta a^{3}L^{2}d_{0}}{5\kappa}\label{eq:polar-act-num}
\end{alignat}
The principal role of the active sphere is to produce both a direct
local force on the terminus of the filament and an indirect non-local
force on the remaining parts of the filament mediated through the
active contribution to the hydrodynamic flow. We now investigate the
actuating dynamics of passive filament in the parameter space defined
by the above two dimensionless groups. 

Fig. (\ref{fig:AtomicFlow}) provides a good approximation to the
local fluid flow near the terminus, with panel (a) corresponding to
$\mathcal{A}^{3t}<0$, and panels (b) and (c) corresponding, respectively,
to $\mathcal{A}^{2s}>0$ and $\mathcal{A}^{2s}<0$. 

\section{Dynamics of actuation\label{sec:Actuating-dynamics}}

As a prelude to presenting our main results, we first study the dynamics
of and active and a passive colloid. The dynamics of the pair is presented
in the figure \ref{fig:AtomicFlow}. In panel (a), the active colloid
is polar and motile with ${\bf V}_{N}^{\mathcal{A}}\neq0$, while
in panels (b) and (c) it is apolar and non-motile with ${\bf V}_{N}^{(2s)}\neq0$.
The choice of the sign of the principal value of $\mathbf{V}_{N}^{(2s)}$
corresponds to extensile (``pusher'') and contractile (``puller'')
forms of apolar activity. The fluid-flow around the assembly is computed
by using the Eq. \ref{eq:flow-comp} and the directions of movement
of the pair for different cases are shown in white arrows. The direction
and the speed of the pair critically depend on the relative position
and orientation configuration of the colloids and the modes of activity.
Surprisingly, we find that even a non-motile active colloid can function
as a propulsion engine in the vicinity of a passive colloid. ``Shakers''
become ``movers'' in ``passive'' company. The significance of
this observation for the colloid-filament assembly is explained below.

We turn now to our main numerical results. In Fig. (\ref{fig:PhaseDiag})
we show the state diagram, in the plane of the two dimensionless activity
parameters, for the filament-colloid assembly. Positive (negative)
values of $\mathcal{A}^{2s}$ corresponds to extensile (contractile)
active flows while positive (negative) values of $\mathcal{A}^{3t}$
corresponds to self-propulsion outwards (towards) the assembly. The
net effect of the active colloid is to produce a force which tends
to extend (compress) the filament, when the parameters are in the
second (fourth) quadrants of the $\mathcal{A}^{2s}-\mathcal{A}^{3t}$
plane. Thus, moving diagonally from the second to the fourth quadrant
leads to an increasing active compression on the filament. In sequence,
we observe a linear state, states in which the filament is non-linear
but has a steady conformation, and finally states in which the filament
conformation is a periodic or aperiodic function of time. The specific
values of the activity numbers at which these states appear depend
on the boundary condition (free, tethered, or clamped) but their sequence
remains unaltered. 

In Fig. (\ref{fig:Flowphase}) we show examples of the linear, non-linear
steady and non-linear unsteady states of actuation, for each kind
of boundary condition and for increasing values of net compressive
activity, with the streamlines of fluid flow superimposed. The locus
of the filament terminus is shown as a solid line for the non-linear
states. The Supplementary Information contains animations of some
of these states.
\begin{figure*}[t]
\includegraphics[width=0.95\textwidth]{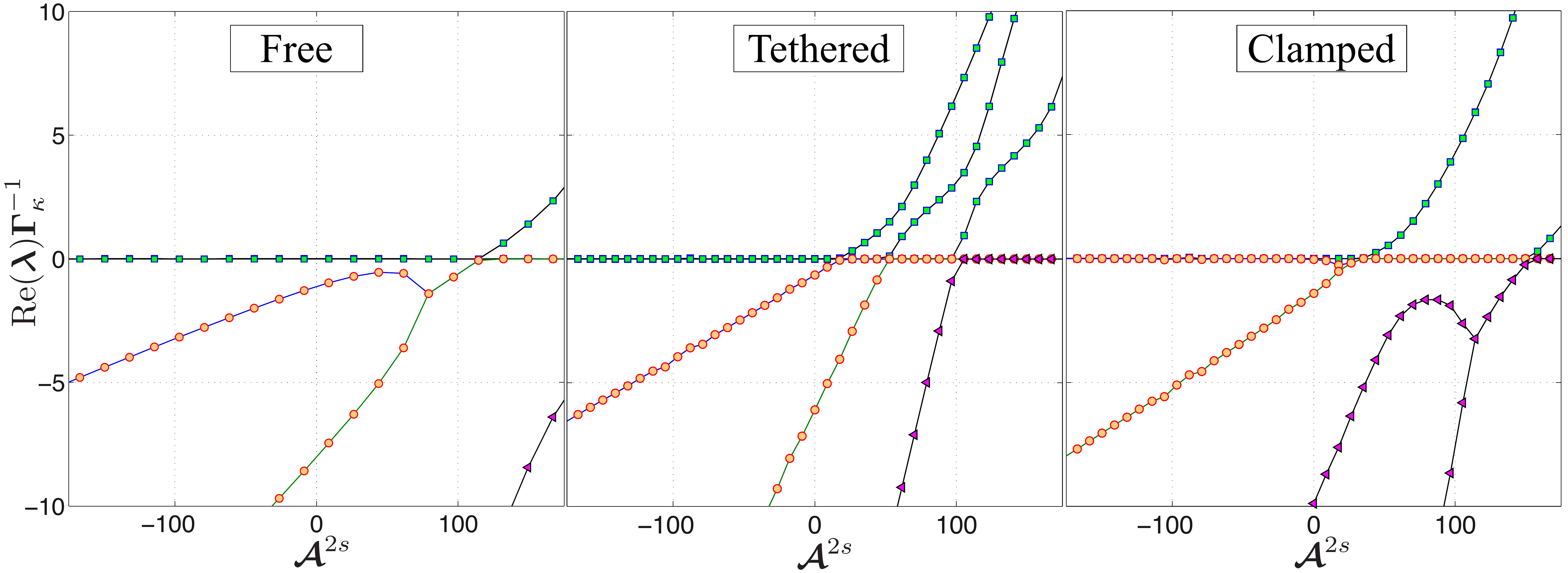}\caption{Variation of the real parts of the largest nonzero eigenvalues $\lambda$
of the stability matrix are plotted against dimensionless measure
of apolar activity, $\mathcal{A}^{2s}$ for free, tethered and clamped
boundary conditions. The transitions from the linear state to non-linear
state proceeds via a Hopf bifurcation for the free and clamped cases
and through a simple instability in the tethered case. $\Gamma_{\kappa}$
is the elastic relaxation rate \cite{jayaraman2012autonomous}. \label{fig:lsa}}
\end{figure*}

The nature of the dynamical transition from the linear state to the
non-linear states can be quantified by a linear stability analysis.
The variation of the largest non-zero eigenvalues of the stability
matrix as a function of $\mathcal{A}^{2s}$ are shown in Fig. (\ref{fig:lsa}).
The transition to non-linear state is through a Hopf bifurcation for
free and clamped filaments but through a simple instability for the
tethered for the filament. This is in contrast to a filament of hydrodynamically
active dipoles in which the free and tethered states have simple instabilities
while the clamped state has a Hopf bifurcation \cite{laskar2013hydrodynamic,laskar2015brownian}.
\begin{figure}[H]
\includegraphics[width=0.95\columnwidth]{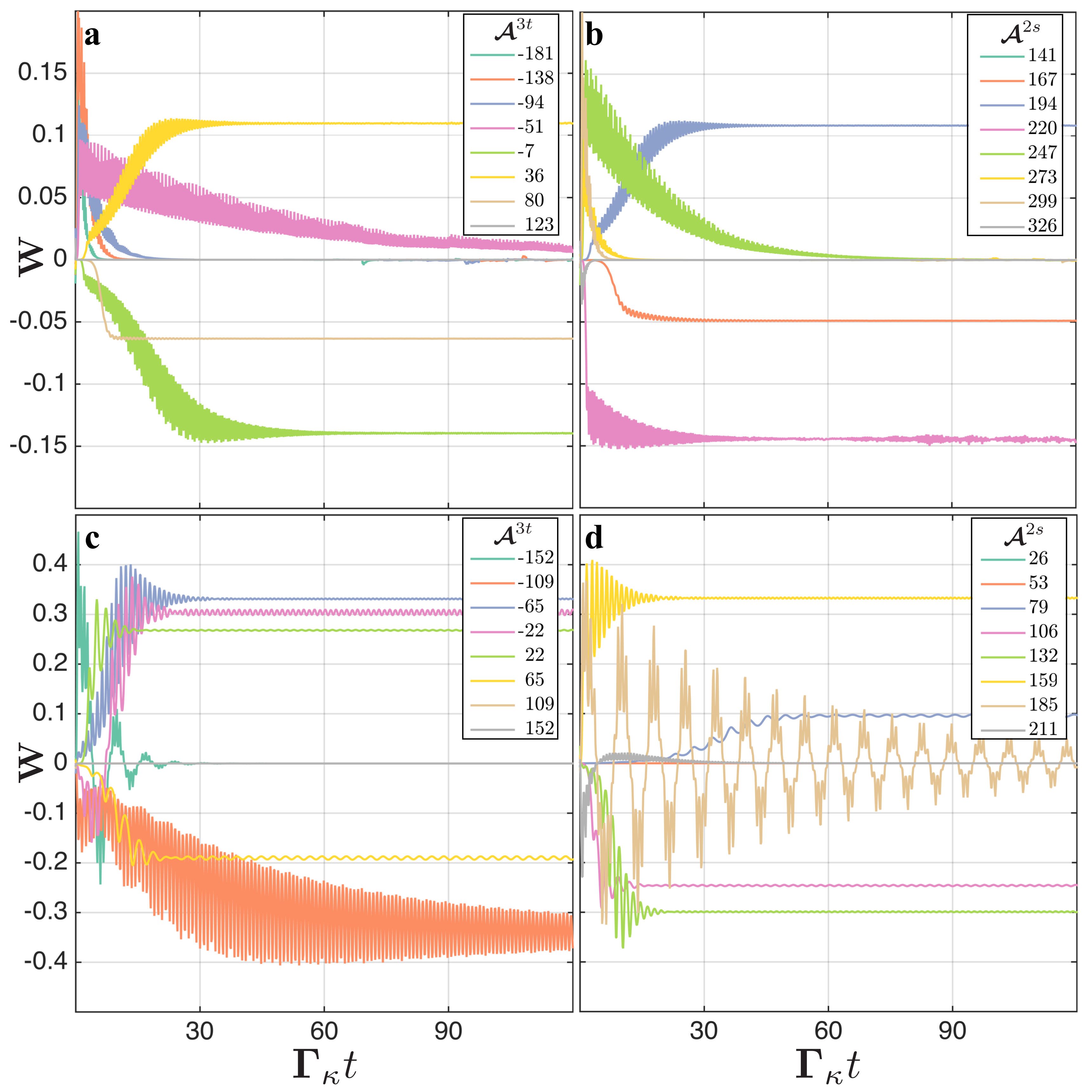}\caption{Time traces of the writhe of the assembly are presented with varying
polar and apolar dimensionless numbers $\mathcal{A}^{3t}$ and $\mathcal{A}^{2s}$
for free (first row) and clamped assembly (second row). For all the
subfigures, we see, that, for low activity both polar and apolar,
the assembly relaxes back to linear conformation; in intermediate
activity region, the assemblies, reach to steady states with constant
or slightly oscillating writhe; at high activity, the assemblies,
goes to a steady state with planar beating with zero writhe. Thus,
using writhe along with the curvature, different regimes of steady
states can clearly be demarcated. \label{fig:writhe}}
\end{figure}

In Fig. (\ref{fig:writhe}) we show the time series of the writhe, 

\begin{equation}
W=\frac{1}{4\pi}\int_{C}\int_{C}d{\bf r}_{1}\times d{\bf r}_{2}\cdot\frac{{\bf r}_{1}-{\bf r}_{2}}{\vert{\bf r}_{1}-{\bf r}_{2}\vert^{3}}\label{eq:writhe-expression}
\end{equation}
a measure of the helicity of a three dimensional curve $C$, where,
${\bf r}_{1}$ and ${\bf r}_{2}$ are points on the curve. We use
the method in \cite{klenin2000computation} to estimate the integral
from the discrete representation of the curve. The writhe and the
curvature are used to demarcate the states in the Fig. (\ref{fig:PhaseDiag}).
In earlier work, the convex hull of the filament conformation was
used as an order parameter but we have found the writhe to be a more
accurate and discriminatory measure of the stationary states. 

We emphasise that the compressive forces that produce the rich dynamics
have both a local contribution, communicated directly at the point
of contact of the filament and the colloid and a non-local contribution
that is mediated by the active flow produced by the colloid. While
the local contribution acts directly on the bead to which the colloid
is attached, the non-local contributions act on all the beads of the
filament, though the strength decreases as the inverse of the square
of the separation between bead and colloid. This non-local transmission
of compressive forces is possible only when momentum conservation
in the fluid is correctly accounted for. Hydrodynamic interactions,
therefore, are of crucial importance in the dynamics uncovered here. 

From the above analysis, it is now possible to choose states for specific
biomimetic applications. The linear state of the free filament is
the most efficient for transport and propulsion. Here, only a very
small part of the active energy input is stored elastically in the
filament as a small compression or extension. The balance is spent
in transport. These, then, are the most efficient states for active
transport. The non-linear states with steady conformations produce
flow fields that co-rotate with the filament. These may be useful
for producing low Reynolds number vortex fields in which large particles
can be trapped or for stirring the medium. The non-linear states with
unsteady conformations produce flow fields that promote efficient
mixing of fluid. These may be used as components of artificial ciliary
carpets that can mix and transport fluid along channels. It is conceivable
that the filament-colloid assembly will find other imaginative uses
in biomimetic applications \cite{singh2015nano,guix2014nano,garcia2013functionalized}.

\section{Discussion\label{sec:Discussions}}

While the actuation mechanism presented here is undoubtedly important
in biomimetic applications it also has a connection to the study of
``follower'' forces in the mechanics of beams. These are forces
that are always directed inwards along the local tangent at the end
of the beam. The analysis of the instabilities of such a beam involves
non-conservative forces for which the classical method of Euler is
inapplicable \cite{bolotin1969effects}. It is has been difficult
to experimentally realize such beams and they remain a controversial
topic in the theory of elastic stability \cite{elishakoff2005controversy}.
Our design points to a simple experimental realization of a slender
elastic body driven by a ``follower'' force exerted by the active
colloid. In addition, it provides yet another interesting angle for
theoretical study, that is, the role of dissipation in elastic instabilities
driven by non-conservative forces. We suggest these are interesting
fields of enquiry in the growing literature on active filaments \cite{chelakkot2014flagellar,jiang2014hydrodynamic,ghosh2014dynamics,jiang2014motion,isele2015self,winkler2016dynamics,isele2016dynamics}.

In this contribution, we have demonstrated that an elastic passive
filament can be actuated without any external field by attaching an
active colloid at its terminus. Though we have developed a theory
considering Brownian motion, their contribution has been ignored in
the present analysis. Here we conclude by pointing that the interplay
between activity and Brownian motion has many interesting consequences
that remain to be explored. 

\section*{Acknowledgement}

The authors thank R. Singh, Arti Dua, R. Manna, and P. B. Sunil Kumar
for many fruitful discussions, the latter two for suggesting the use
of writhe as an order parameter, The Institute of Mathematical Sciences
for providing access to computing resources on the Annapurna and Nandadevi
clusters, and the Department of Atomic Energy, Government of India
for supporting their research.

\appendix

\section{Hydrodynamic tensors and fluid flow}

The mobility and the propulsion tensors can be computed to any desired
accuracy and order \cite{singh2016traction}. The leading order forms
in an unbounded medium, where $G_{ij}=\frac{\delta_{ij}}{r}+\frac{r_{i}r_{j}}{r^{3}}$
, are 

\begin{alignat}{1}
8\pi\eta\bm{\mu}_{nm}^{TT} & =\begin{cases}
\frac{4}{3a}\bm{\delta} & \qquad\qquad\negthickspace m=n\\
\mathcal{F}^{0}\mathcal{F}^{0}\mathbf{G}(\mathbf{R}_{n},\mathbf{R}_{m}) & \qquad\qquad\negthickspace m\neq n
\end{cases}\\
8\pi\eta\bm{\mu}_{nm}^{TR} & =\begin{cases}
0 & \qquad\quad\negthickspace m=n\\
\frac{1}{2}\bm{\nabla}_{m}\times\mathbf{G}(\mathbf{R}_{n},\mathbf{R}_{m}) & \qquad\quad\negthickspace m\neq n
\end{cases}\\
8\pi\eta\bm{\mu}_{nm}^{RT} & =\begin{cases}
0 & \qquad\quad\negthickspace m=n\\
\frac{1}{2}\bm{\nabla}_{n}\times\mathbf{G}(\mathbf{R}_{n},\mathbf{R}_{m}) & \qquad\quad\negthickspace m\neq n
\end{cases}\\
8\pi\eta\bm{\mu}_{nm}^{RR} & =\begin{cases}
\frac{1}{a}\bm{\delta} & \negthickspace m=n\\
\frac{1}{4}\bm{\nabla}_{n}\times\bm{\nabla}_{m}\times\mathbf{G}(\mathbf{R}_{n},\mathbf{R}_{m}) & \negthickspace m\neq n
\end{cases}
\end{alignat}

The diagonal parts of these matrices are one-body terms while the
off-diagonal parts represent the hydrodynamic interactions. The diagonal
parts are the familiar Stokes translational and rotational mobilities
while the off-diagonal parts can be recognised as the Rotne-Prager-Yamakawa
tensors \cite{rotne1969variational,yamakawa1970transport} and their
generalizations to rotational motion. The Onsager symmetry of the
mobility matrix is manifest in these expressions. Similarly, the propulsion
matrices can also be computed \cite{singh2014many}, which are 

\begin{alignat}{1}
\bm{\pi}_{nm}^{(T,\,l)} & =\begin{cases}
\bm{\delta} & m=n,\,l\sigma=1s\\
0 & m=n,\,l\sigma\neq1s\\
c_{l}\mathcal{F}^{0}\mathcal{F}^{(l-1)}\mathbf{\bm{\nabla}}_{m}^{(l-1)}\mathbf{G}(\mathbf{R}_{n},\mathbf{R}_{m}) & m\neq n
\end{cases}\\
\bm{\pi}_{nm}^{(R,\,l)} & =\begin{cases}
\frac{1}{a}\bm{\delta} & \quad\:m=n,\,l\sigma=2a\\
0 & \quad\:m=n,\,l\sigma\neq2a\\
\dfrac{c_{l}}{2}\bm{\nabla}_{n}\times\mathbf{\bm{\nabla}}_{m}^{(l-1)}\mathbf{G}(\mathbf{R}_{n},\mathbf{R}_{m}) & \quad\:m\neq n
\end{cases}
\end{alignat}

The form of mobility and propulsion matrices, we have thus got for
beads of radius $a$ by considering solution after the first iteration,
can be computed through alternative way through the pair-wise superposition
approximation, first introduced by Kirkwood and Riseman \cite{kirkwood1948intrinsic},
in their contributions on the dynamics of a polymer.

The fluid flow exterior to the filament-colloid assembly, with the
two-mode truncation for the slip, is 
\begin{align}
\mathbf{v}(\mathbf{r})=-\frac{1}{8\pi\eta}\sum_{m=1}^{N}\mathcal{F}^{0}\mathbf{G}\cdot\bm{\nabla}_{m}U & +\overbrace{\frac{7a^{2}}{6}\mathcal{F}^{1}\bm{\nabla}_{N}\mathbf{G}\cdot{\bf V}_{N}^{(2s)}}^{\mathrm{active}}\nonumber \\
 & +\frac{a^{3}}{10}\nabla_{N}^{2}\mathbf{G}\cdot{\bf V}_{N}^{(3t)}.\label{eq:flow-comp}
\end{align}
This expression is used to plot the streamlines in the Fig.(\ref{fig:AtomicFlow})
and Fig.(\ref{fig:Flowphase}).

\section{Simulation parameters}

The parameters we choose for the simulation as follows: bond-length
$b_{0}=4a$, bending rigidity $\kappa=0.1$, spring constant $k=1$,
non-motile activity or stresslet strength $s_{0}=-0.5-0.5$ and motile
activity or degenerate quadrupole strength $d_{0}=-0.005-0.005$.
The number of beads in the filament is $N=32$. We simulate the system
for several hundred passive relaxation times $\Gamma_{\kappa}^{-1}$
while computing the hydrodynamic tensors at each time step using the
PyStokes library \cite{singh2014pystokes}. The initial condition
in all simulations is the linear state with small-amplitude random
transverse perturbations.
\begin{table}[H]
\renewcommand{\arraystretch}{1.5} 

\begin{tabular}[b]{|c|c|c|c|}
\hline 
\multirow{2}{*}{} & \multirow{2}{2cm}{Linear} & \multirow{2}{2cm}{Rotation or Helical} & \multirow{2}{2cm}{Planar mode}\tabularnewline
 &  &  & \tabularnewline
\hline 
\multirow{2}{*}{Free} & $\mathcal{A}^{2s}\simeq70$ & $\mathcal{A}^{2s}\simeq175$ & $\mathcal{A}^{2s}\simeq245$\tabularnewline
 & $\mathcal{A}^{3t}\simeq90$ & $\mathcal{A}^{3t}\simeq0$ & $\mathcal{A}^{3t}\simeq-60$\tabularnewline
\hline 
\multirow{2}{*}{Tethered} & $\mathcal{A}^{2s}\simeq0$ & $\mathcal{A}^{2s}\simeq70$ & $\mathcal{A}^{2s}\simeq140$\tabularnewline
 & $\mathcal{A}^{3t}\simeq90$ & $\mathcal{A}^{3t}\simeq60$ & $\mathcal{A}^{3t}\simeq-30$\tabularnewline
\hline 
\multirow{2}{*}{Clamped} & $\mathcal{A}^{2s}\simeq-35$ & $\mathcal{A}^{2s}\simeq135$ & $\mathcal{A}^{2s}\simeq210$\tabularnewline
 & $\mathcal{A}^{3t}\simeq120$ & $\mathcal{A}^{3t}\simeq0$ & $\mathcal{A}^{3t}\simeq-90$\tabularnewline
\hline 
\end{tabular}

\caption{Activity numbers correspond to panels of \ref{fig:Flowphase}.}
\end{table}

\bibliographystyle{unsrt}

\end{document}